# Single-Antenna-Based GPS Antijamming Method Exploiting Polarization Diversity

Kwansik Park and Jiwon Seo, *Member, IEEE*

*Abstract*—The vulnerability of Global Positioning System (GPS) receivers to jammers is a major concern owing to the extremely weak received signal power of GPS. Researches have been conducted on a variety of antenna array techniques to be used as countermeasures to GPS jammers, and their antijamming performance is known to be greater than that of single antenna methods. However, the application of antenna arrays remains limited because of their size, cost, and computational complexity. This study proposes and experimentally validates a novel space-time-polarization domain adaptive processing for a single-element dual-polarized antenna (STPAPS) by focusing on the polarization diversity of a dual-polarized antenna. The mathematical models of arbitrarily polarized signals received by dual-polarized antenna are derived, and an appropriate constraint matrix for dual-polarized-antenna-based GPS antijam is suggested. To reduce the computational complexity of the constraint matrix approach, the eigenvector constraint design scheme is adopted. The performance of STPAPS is quantitively and qualitatively evaluated through experiments as follows. 1) The carrier-to-noise-density ratio ($C/N_0$) of STPAPS under synthetic jamming is demonstrated to be higher than that of the previous minimum mean squared error (MMSE) or minimum variance distortionless response (MVDR) based dual-polarized antenna methods. 2) The strengths and weaknesses of STPAPS are qualitatively compared with those of the previous single-element dual-polarized antenna methods that are not based on the MMSE or MVDR algorithms. 3) The characteristics of STPAPS (in terms of the directions and polarizations of the GPS and jamming signals) are compared with those of the conventional two-element single-polarized antenna array method, which has the same degree of freedom as that of STPAPS. This is a GPS L1 C/A-code proof-of-concept study, although STPAPS will extend to higher chipping-rate modernized global navigation satellite system (GNSS) signals.

*Index Terms*—Global Positioning System (GPS), antijamming, polarization diversity, dual-polarized antenna

Manuscript received;

This research was supported by the Unmanned Vehicles Core Technology Research and Development Program through the National Research Foundation of Korea (NRF) and the Unmanned Vehicle Advanced Research Center (UVARC) funded by the Ministry of Science and ICT, Republic of Korea (No. 2020M3C1C1A01086407). This work was also supported by the Institute for Information & Communications Technology Planning & Evaluation (IITP) grant funded by the Korea government (KNPA) (No. 2019-0-01291).

K. Park and J. Seo are with the School of Integrated Technology, College of Engineering, Yonsei University, 85 Songdogwahak-ro, Yeonsu-gu, Incheon 21983, Korea. (e-mail: jiwon.seo@yonsei.ac.kr).

## I. INTRODUCTION

ALTHOUGH the Global Positioning System (GPS) plays a key role in numerous civilian and military applications [1]–[7] including aviation [8]–[12], their vulnerability to jammers remains a major concern [13]–[15]. Because the received GPS signal power is extremely low, typically more than 20 dB below the noise floor [16], even very low-power jamming signals can easily overwhelm the GPS signals.

As countermeasures to GPS jamming signals, antenna-array-based signal processing methods have been investigated intensively, and have shown better antijamming performance than conventional single-antenna-based techniques. A typical $N$-element GPS antenna array has $N-1$ spatial degrees of freedom (DOFs). Thus, by spatial processing, it can steer up to $N-1$ spatial nulls (i.e., low signal processing gains in certain directions) toward jammers and thereby mitigate $N-1$ narrowband jamming signals. (In this study, the term narrowband signal denotes a signal whose ratio of bandwidth to center frequency is less than 2% [17]. Thus, GPS signals are narrowband signals.) A GPS controlled reception pattern antenna (CRPA) [18]–[21] is typically implemented using an antenna array and finite impulse response (FIR) filters attached to the antenna elements. In this way, the frequency notches (i.e., low signal processing gains at certain frequencies) formed by multiple filter time taps can also suppress additional continuous-wave (CW) jammers [19]. Therefore, the total number of jamming signals that can be mitigated may exceed the spatial DOFs of the array if CW jammers are present. The DOFs of the array can be increased by using a dual-polarized array [22].

Even though antenna-array-based techniques have greater jammer mitigation capabilities than single-antenna-based techniques, antenna-array methods have limited applications due to the large array size, high cost, and computational complexity. Therefore, the approach proposed in this study, namely space-time-polarization domain adaptive processing for a single-element dual-polarized antenna (STPAPS), involves focusing on the polarization diversity of a single-element dual-polarized antenna to reduce the size of the solution. STPAPS utilizes the minimum variance distortionless response (MVDR) beamformer [23]. Although MVDR-based spatial-temporal adaptive processing (STAP) for GPS CRPA is well known [18], [19], [24], [25], the optimal weight vector for the conventional single-polarized GPS CRPA does not provide jammer mitigation capability to a dual-polarized antenna. Therefore, STPAPS expands MVDR-based STAP methods to include polarization domain processing, which makes it applicable to a single-element dual-polarized antenna.

There exist other dual-polarized antenna-based GPS

antijamming methods in the literature. Fante [26] and Park et al. [22] proposed minimum mean squared error (MMSE) and MVDR based antijamming algorithms for dual-polarized antenna arrays, respectively. Although these methods were originally developed for antenna arrays, they are applicable to single-element dual-polarized antennas as well by treating a single-element antenna as a one-element array. Wang et al. [27] utilized the constraint vector proposed by Park et al. [22] for their implementation and suggested a method to improve its computational efficiency. In addition to these efforts, we presented the possibility to expand the existing spatial adaptive processing (SAP) or STAP algorithms to include polarization-domain processing for a dual-polarized antenna in [28], [29]. However, their performance was not very satisfactory. The method in [28] can mitigate only a single jammer, and the method in [29] undergoes performance degradation if the incident jammer and GPS signal are correlated.

Unlike the MMSE algorithm, constraint design is very important for the MVDR algorithm. The aforementioned studies regarding MVDR algorithms used a single constraint vector corresponding to a specific look direction, and the constraint vector constrains the weights corresponding to only a single time tap of the FIR filters that are used to implement STAP. However, GPS signals are not single tone but have a spectrum spread over a frequency band. In addition, when STAP is implemented using an $N$-element antenna array and the attached FIR filters with $M$ time taps, a total of $NM$ weights is involved in the processing gain calculation. Nevertheless, the constraint vectors in the above-mentioned previous works consist of the first $N$ non-zero terms and the remaining $N(M-1)$ zeros. Since the $N(M-1)$ weights corresponding to those zeros are not constrained, undesired attenuation of GPS signals can occur during jammer mitigation.

For these reasons, STPAPS adopts a constraint matrix approach that consists of constraint vectors corresponding to multiple frequency points within a frequency band of interest, rather than a single constraint vector as used in previous works [18], [19], [24], [25]. Each constraint vector of the constraint matrix is derived from our received signal model of the single-element dual-polarized antenna and is designed to constrain all the weights corresponding to every filter time tap with respect to the direction and polarization of the GPS signal. In addition, to avoid increasing the computational complexity due to numerous constraint vectors, the eigenvector constraint design [30]–[34] based on low-rank representation is employed. This design approach makes it possible to find an approximate constraint matrix with a relatively small number of constraint vectors and minimize the difference between the desired and actual signal processing gains over the frequency band of interest. Finally, the optimal weight vector for the antijamming operation of each tracking channel of the receiver is calculated by using the approximate constraint matrix, without the need for any prior knowledge of the jammer polarizations and directions.

Other methods that are not based on the MMSE or MVDR algorithms have also been developed specifically for single-element dual-polarized antennas. They are also compared with STPAPS in this study. Rosen and Braasch [35] proposed an antijamming technique based on the observed polarization difference between GPS and jamming signals, which is effective for mitigating a jammer that is not right-hand circularly polarized (RHCP). This method can even mitigate multiple wideband jamming signals if the jammer polarizations are identical. The methods of Kraus et al. [36] and McMilin et al. [37] can mitigate a single jamming signal by adaptively shifting the phase of a signal from one port of the antenna and then combining the phase-shifted signal with the signal from the other port. In contrast with the method in [36], which can suppress only a linearly polarized (LP) jamming signal, the technique in [37] is capable of mitigating a jammer with any polarization originating from below the horizon, thereby making it suitable for aviation applications.

The performance and characteristics of STPAPS are evaluated as follows. 1) The carrier-to-noise-density ratio ($C/N_0$) of STPAPS under synthetic jamming is quantitively compared with that of the other dual-polarized antenna techniques [22], [26], which utilize the MMSE or MVDR algorithms, in both low and high multipath environments. 2) The strengths and weaknesses of STPAPS are qualitatively discussed in comparison with the other single-element dual-polarized antenna methods [35]-[37] that are not based on the MMSE or MVDR algorithms. Quantitative evaluation was not possible in this case because the methods in [35], [37] used specially designed antenna hardware that were not available to us, and the exact algorithm of [36] is not revealed in the paper. 3) The characteristics of STPAPS and the conventional two-element single-polarized array STAP are compared in terms of the directions and polarizations of GPS and jamming signals. Both methods have the same DOF, and therefore it is worth comparing; however, the characteristics of STAP and space-time-polarization domain processing techniques such as STPAPS have not been compared well in the literature.

Under the objective of providing a small form-factor GPS antijamming solution while maintaining or even improving its antijamming performance compared with that of existing solutions, this study makes the following contributions: To overcome the limitation of the existing method, which relies on a single-frequency constraint vector, we adopted the constraint matrix approach and derived an appropriate constraint matrix for a single-element dual-polarized antenna. To increase its computational efficiency, the eigenvector constraint design scheme was first adopted for the purpose of GPS antijam. To demonstrate the performance of STPAPS, experiments were performed in both low and high multipath environments, and the performance of STPAPS was compared with that of other antijamming methods in three different categories as explained in the previous paragraph.

The remainder of this paper is organized as follows. In Section II, the mathematical models of the signals received by a dual-polarized antenna are derived. Section III proposes STPAPS by utilizing the eigenvector constraint design scheme. In Section IV, the antijamming performance and characteristics of STPAPS are discussed based on the experimental results. Section V presents the conclusions.

## II. Problem Formulation

Let us consider a dual-polarized antenna located at the origin of the coordinate system shown in Fig. 1, where $\hat{\mathbf{e}}_\varphi$, $\hat{\mathbf{e}}_\theta$, and $\hat{\mathbf{e}}_r$ are the respective unit vectors of the spherical coordinates. The dual-polarized antenna used in this work consists of RHCP and left-hand circularly polarized (LHCP) antenna components.

### A. Mathematical model of incoming polarized signals

In general, an electromagnetic (EM) plane wave has an arbitrary elliptical polarization. The polarization state of an EM wave traveling in the $-\hat{\mathbf{e}}_r$ direction can be determined by the relative magnitudes and phases of two perpendicular electric field components. The electric field of the incoming signal with amplitude $E_0$ in the complex vector form can be represented as follows [38]:

$$\mathbf{E}(\varphi,\theta,\gamma,\eta) = E_0(\cos\gamma\,\hat{\mathbf{e}}_\varphi + \sin\gamma\exp[j\eta]\hat{\mathbf{e}}_\theta) \\ = E_0\hat{\mathbf{E}}(\varphi,\theta,\gamma,\eta) \tag{1}$$

where $\gamma$ ($0 \leq \gamma \leq \pi/2$) and $\eta$ ($-\pi \leq \eta \leq \pi$) are angular quantities used for describing the signal polarization and $\hat{\mathbf{E}}(\varphi,\theta,\gamma,\eta)$ is the normalized electric field vector of unit magnitude. The values of $\gamma$ and $\eta$ uniquely determine the polarization state of the signal.

### B. Mathematical model of signals received by a dual-polarized antenna

When an EM wave impinges on an antenna, a voltage is induced across the antenna. The voltage is proportional to the inner product between the complex radiation field function of the receiving antenna and the electric field vector of the incoming wave, as follows [39]:

$$V = K\mathbf{G}(\varphi,\theta,f) \cdot \mathbf{E}(\varphi,\theta,\gamma,\eta)^* \\ = \left\{K\mathbf{G}(\varphi,\theta,f) \cdot \hat{\mathbf{E}}(\varphi,\theta,\gamma,\eta)^*\right\} E_0 \tag{2}$$

where $*$ is the complex conjugate operator. The scaling factor $K$ is independent of the signal direction and polarization but dependent on other parameters such as the intrinsic impedance of the propagation medium. The vector $\mathbf{G}(\varphi,\theta,f)$ is the complex radiation field function of the antenna corresponding to the direction ($\varphi$, $\theta$) and frequency $f$, and $\mathbf{G}(\varphi,\theta,f)$ consists of the gain and phase patterns corresponding to two orthogonal directions, as follows:

$$\mathbf{G}(\varphi,\theta,f) = G_\varphi(\varphi,\theta,f)\exp\left[jP_\varphi(\varphi,\theta,f)\right]\hat{\mathbf{e}}_\varphi \\ + G_\theta(\varphi,\theta,f)\exp\left[jP_\theta(\varphi,\theta,f)\right]\hat{\mathbf{e}}_\theta \tag{3}$$

where $G_\varphi(\varphi,\theta,f)$ and $G_\theta(\varphi,\theta,f)$ are the gain patterns and $P_\varphi(\varphi,\theta,f)$ and $P_\theta(\varphi,\theta,f)$ are the phase patterns of the receiving antenna in the direction of ($\varphi$, $\theta$), corresponding to the incoming signals that have frequency $f$ and are polarized along the directions of $\hat{\mathbf{e}}_\varphi$ and $\hat{\mathbf{e}}_\theta$, respectively. Even though the complex radiation field function $\mathbf{G}(\varphi,\theta,f)$ depends on the signal frequency, antenna patterns with respect to the center frequency of the GPS signal have been used as representative patterns in the previous studies [40], [41]. For this reason, the notation $\mathbf{G}(\varphi,\theta)$ is used instead of $\mathbf{G}(\varphi,\theta,f)$ in the remainder of this paper.

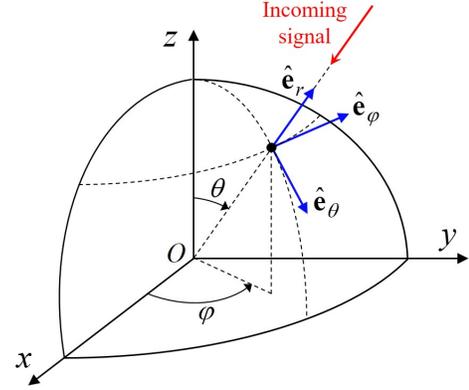

Fig. 1. Coordinate systems for the dual-polarized antenna.

Because the dual-polarized antenna in this study consists of RHCP and LHCP antenna components, the two voltages, $V_R$ and $V_L$, can be obtained from the two antenna components, with respect to the incoming signal. The use of (2) enables us to express these two voltages as follows:

$$V_R = \left\{K\mathbf{G}_R(\varphi,\theta) \cdot \hat{\mathbf{E}}(\varphi,\theta,\gamma,\eta)^*\right\} E_0 \\ V_L = \left\{K\mathbf{G}_L(\varphi,\theta) \cdot \hat{\mathbf{E}}(\varphi,\theta,\gamma,\eta)^*\right\} E_0 \tag{4}$$

where $\mathbf{G}_R(\varphi,\theta)$ and $\mathbf{G}_L(\varphi,\theta)$ are the complex radiation field functions of the RHCP and LHCP antenna components, respectively. In the case of conventional antenna arrays, the signals received by the antenna elements of the array differ in their relative phases, and the relative phases are dependent on the direction of the incoming signal. These phase differences provide spatial DOFs and enable the antenna array to distinguish the jammers from the GPS signal in the spatial domain. On the other hand, in the case of the single-element dual-polarized antenna, the signals received by the RHCP and LHCP antenna components undergo different amounts of amplitude and phase changes, as shown in (4). This difference provides a DOF for jammer mitigation. In Section III, we propose an adaptive antijamming algorithm based on the signal model derived in this section.

## III. Adaptive Signal Processing Method for a Dual-Polarized Antenna

The concept of STPAPS, which is proposed in this paper, is illustrated in Fig. 2. The two antenna components are followed by FIR filters with $M$ time taps. After down-conversion from radio frequency to intermediate frequency (IF) and sampling, at the $k$-th sampling epoch, the complex IF output signal $u[k]$ is represented as follows:

$$u[k] = \sum_{m=0}^{M-1}\left\{w_{R,m}^*v_{R,m}[k] + w_{L,m}^*v_{L,m}[k]\right\} \\ = \mathbf{w}_{2M\times 1}^H\mathbf{v}_{2M\times 1}[k] \tag{5}$$

where:

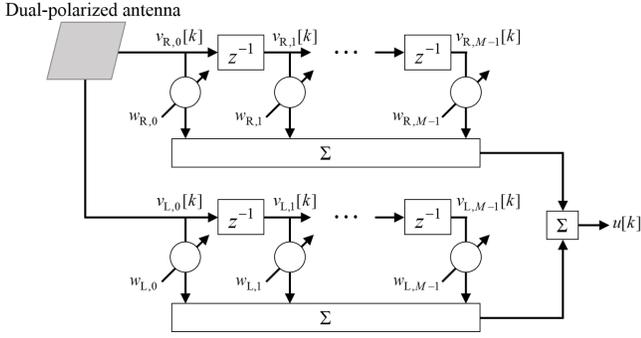

Fig. 2. Block diagram of STPAPS. The signals from the RHCP and LHCP feeds of the single-element dual-polarized antenna are processed and combined to mitigate jammers.

$$\mathbf{w} = \begin{bmatrix} w_{R,0} & w_{L,0} & \cdots & w_{R,M-1} & w_{L,M-1} \end{bmatrix}^T$$

$$\mathbf{v}[k] = \begin{bmatrix} v_{R,0}[k] & v_{L,0}[k] & \cdots & v_{R,M-1}[k] & v_{L,M-1}[k] \end{bmatrix}^T \quad (6)$$

Here, $\mathbf{w}$ is a complex weight vector, whose elements $w_{R,m}$ and $w_{L,m}$ are multiplied by the signals $v_{R,m}[k]$ and $v_{L,m}[k]$ at the $m$-th time taps of the FIR filters attached to the two antenna components, and $\mathbf{v}[k]$ is the complex IF input signal vector. (The number of delayed samples for each filter tap is one.).

### A. Signal processing gain of STPAPS

By using the received signal model in (4) and the relationship between the received signals and the output signal in (5), the signal processing gain of STPAPS (shown in Fig. 2) can be obtained. Let us consider an incoming signal, $s(t)$, for which the direction, polarization, and normalized electric field are specified by $(\varphi, \theta)$, $(\gamma, \eta)$, and $\hat{\mathbf{E}}(\varphi, \theta, \gamma, \eta)$, respectively. Using (4) and (5), in the absence of any other signals and noise, the output signal can be represented as follows:

$$u[k] = K \sum_{m=0}^{M-1} \left\{ w_{R,m}^* \mathbf{G}_R(\varphi,\theta) \cdot \hat{\mathbf{E}}(\varphi,\theta,\gamma,\eta)^* + w_{L,m}^* \mathbf{G}_L(\varphi,\theta) \cdot \hat{\mathbf{E}}(\varphi,\theta,\gamma,\eta)^* \right\} s[k-m] \quad (7)$$

where $s[k]$ is the down-converted and sampled version of $s(t)$. The Fourier transform of (7) is expressed as follows:

$$U(f) = S(f) K \left[ \sum_{m=0}^{M-1} \left\{ w_{R,m}^* \mathbf{G}_R(\varphi,\theta) \cdot \hat{\mathbf{E}}(\varphi,\theta,\gamma,\eta)^* + w_{L,m}^* \mathbf{G}_L(\varphi,\theta) \cdot \hat{\mathbf{E}}(\varphi,\theta,\gamma,\eta)^* \right\} \cdot \exp\left[-j\frac{2\pi m f}{f_s}\right] \right] \quad (8)$$

where $S(f)$ and $U(f)$ are the Fourier transforms of $s[k]$ and $u[k]$, respectively, and $f_s$ is the sampling frequency. The ratio of $U(f)$ and $S(f)$ is given by:

$$\frac{U(f)}{S(f)} = K \left[ \sum_{m=0}^{M-1} \left\{ w_{R,m}^* \mathbf{G}_R(\varphi,\theta) \cdot \hat{\mathbf{E}}(\varphi,\theta,\gamma,\eta)^* + w_{L,m}^* \mathbf{G}_L(\varphi,\theta) \cdot \hat{\mathbf{E}}(\varphi,\theta,\gamma,\eta)^* \right\} \cdot \exp\left[-j\frac{2\pi m f}{f_s}\right] \right] \quad (9)$$

Since the scaling factor $K$ is constant, the term in the square brackets on the right-hand side represents the variations of the signal-processing gain of STPAPS, with respect to the incoming signal specified by its four angular quantities ($\varphi$, $\theta$, $\gamma$, $\eta$) and $f$, as follows:

$$g(\varphi,\theta,\gamma,\eta,f) = \sum_{m=0}^{M-1} \left\{ w_{R,m}^* \mathbf{G}_R(\varphi,\theta) \cdot \hat{\mathbf{E}}(\varphi,\theta,\gamma,\eta)^* + w_{L,m}^* \mathbf{G}_L(\varphi,\theta) \cdot \hat{\mathbf{E}}(\varphi,\theta,\gamma,\eta)^* \right\} \cdot \exp\left[-j\frac{2\pi m f}{f_s}\right]$$

$$= \mathbf{w}^H \tilde{\mathbf{c}}_{2M \times 1}(\varphi,\theta,\gamma,\eta,f) \quad (10)$$

where:

$$\tilde{\mathbf{c}}(\varphi,\theta,\gamma,\eta,f) = \begin{bmatrix} \tilde{c}_{R,0} & \tilde{c}_{L,0} & \cdots & \tilde{c}_{R,M-1} & \tilde{c}_{L,M-1} \end{bmatrix}^T \quad (11)$$

and

$$\tilde{c}_{R,m} = \mathbf{G}_R(\varphi,\theta) \cdot \hat{\mathbf{E}}(\varphi,\theta,\gamma,\eta)^* \exp\left[-j\frac{2\pi m f}{f_s}\right]$$

$$\tilde{c}_{L,m} = \mathbf{G}_L(\varphi,\theta) \cdot \hat{\mathbf{E}}(\varphi,\theta,\gamma,\eta)^* \exp\left[-j\frac{2\pi m f}{f_s}\right] \quad (12)$$

As shown in (10), the signal processing gain is dependent on the polarization ($\gamma$, $\eta$) as well as the direction ($\varphi$, $\theta$) and frequency $f$ of the received signal. Due to this polarization dependency, STPAPS can mitigate a non-RHCP jammer while preserving the GPS signal even when the two signals are incident from the same direction. This phenomenon cannot be observed in the case of an $N$-element single-polarized antenna array since the signal processing gain depends only on the signal direction ($\varphi$, $\theta$) and frequency $f$ in the following manner [19]:

$$\bar{g}(\varphi,\theta,f) = \sum_{n=1}^{N} \sum_{m=0}^{M-1} w_{n,m}^* \exp\left[-j\left(\Delta\psi_{\text{GPS},n} + \frac{2\pi m f}{f_s}\right)\right] \quad (13)$$

where $\Delta\psi_{\text{GPS},n}$ ($n = 1, 2, \ldots, N$) is the phase of the GPS signal received by the $n$-th antenna element with respect to the reference antenna element, and it is dependent on both the array geometry and the GPS signal direction. Since all the physical antenna elements of the array are identical, their inner product terms between the complex radiation field functions and normalized electric field vectors are also identical. Therefore, this constant inner product term is not included in (13), as the constant $K$ in (9) is not included in (10), because (10) and (13) express only the variations of the signal processing gain.

The signal processing gain of the proposed method in (10) is expressed by the vector $\tilde{\mathbf{c}}(\varphi,\theta,\gamma,\eta,f)$ corresponding to the direction ($\varphi$, $\theta$), polarization ($\gamma$, $\eta$), and frequency $f$ of the received signal. The weight vector $\mathbf{w}$ in (10) is obtained based on $\tilde{\mathbf{c}}(\varphi,\theta,\gamma,\eta,f)$ which will be explained in Section III-C. Therefore, $\tilde{\mathbf{c}}(\varphi,\theta,\gamma,\eta,f)$ is the key component of STPAPS, and it is proposed by deriving the signal processing gain of a single-element dual-polarized antenna as explained in this subsection.

*B. Conventional constraint vectors of MVDR algorithms for single- or dual-polarized antenna arrays*

In the literature regarding GPS antijam using single- or dual-polarized antenna arrays, a linearly constrained MVDR beamformer is often used to obtain the weight vector to mitigate jammers as follows [42]:

$$\min_{\mathbf{w}} \mathbf{w}^H \mathbf{R} \mathbf{w} \quad \text{subject to} \quad \mathbf{w}^H \mathbf{c}_{NM \times 1} = 1 \quad (14)$$

where $\mathbf{R} = E\{\mathbf{v}[k]\mathbf{v}^H[k]\}$ is the covariance matrix of the received signals, and $\mathbf{c}$ is the constraint vector corresponding to a specific frequency.

In previous works regarding the $N$-element dual-polarized antenna arrays [22], [27], the constraint vector is designed with respect to both the center frequency and the polarization of the received GPS signal, and it constrains the weight corresponding to a single time tap of each FIR filter in the following manner:

$$\mathbf{c} = \begin{bmatrix} \overline{\mathbf{c}}_{1 \times 2N} & \mathbf{0}_{1 \times 2N(M-1)} \end{bmatrix}^T \quad (15)$$

where $\overline{\mathbf{c}}$ given as follows:

$$\overline{\mathbf{c}} = \begin{bmatrix} c_{R,1} & c_{L,1} & \cdots & c_{R,N} & c_{L,N} \end{bmatrix} \quad (16)$$

where

$$\begin{aligned} c_{R,n} &= \mathbf{G}_R(\varphi_{GPS}, \theta_{GPS}) \cdot \hat{\mathbf{E}}(\varphi_{GPS}, \theta_{GPS}, 45°, -90°)^* \\ &\quad \cdot \exp[-j\Delta\psi_{GPS,n}] \\ c_{L,n} &= \mathbf{G}_L(\varphi_{GPS}, \theta_{GPS}) \cdot \hat{\mathbf{E}}(\varphi_{GPS}, \theta_{GPS}, 45°, -90°)^* \\ &\quad \cdot \exp[-j\Delta\psi_{GPS,n}] \end{aligned} \quad (17)$$

On the other hand, in the case of the $N$-element single-polarized antenna arrays studied in [18], [19], [24], [25], the constraint vector is designed as follows:

$$\mathbf{c} = \begin{bmatrix} \overline{\mathbf{c}}_{1 \times N} & \mathbf{0}_{1 \times N(M-1)} \end{bmatrix}^T \quad (18)$$

where

$$\overline{\mathbf{c}} = \begin{bmatrix} \exp[-j\Delta\psi_{GPS,1}] & \cdots & \exp[-j\Delta\psi_{GPS,N}] \end{bmatrix} \quad (19)$$

Even though MVDR-based single- or dual-polarized antenna array algorithms with the constraint vectors given by (15) or (18) have been shown to be effective for jammer mitigation in the previous studies, undesired attenuation of GPS signals can occur. This is because the signal processing gain actually depends on the weights of all the time taps, not just a single tap, and the GPS signal is not single-tone but has a spectrum spread over a frequency band. For these reasons, the usage of a constraint vector synthesized at a single frequency as shown in (15) or (18) can lead to undesired degradation in the performance.

*C. Proposed STPAPS algorithm utilizing constraint matrix and eigenvector constraint design scheme*

Because of the weakness of the conventional algorithms based on a single-frequency constraint vector discussed in the previous subsection, we propose a constraint matrix approach for GPS antijam with a dual-polarized antenna. The constraint matrix consists of multiple constraint vectors for the multiple frequency points in the GPS frequency band, and each constraint vector for a given frequency point constrains the weights for all the filter time taps (i.e., a total of $2M$ time taps in Fig. 2). Given $\varepsilon$ linear constraints, the expression of the MVDR beamformer with the constraint matrix $\mathbf{C}$ can be represented as follows [43]:

$$\min_{\mathbf{w}} \mathbf{w}^H \mathbf{R} \mathbf{w} \quad \text{subject to} \quad \mathbf{w}^H \mathbf{C} = \mathbf{1}_{\varepsilon} \quad (20)$$

where $\mathbf{1}_{\varepsilon}$ is an $\varepsilon$-dimensional row vector whose elements are all equal to one, and $\mathbf{C}$ is a constraint matrix whose columns are $\varepsilon$ constraint vectors that correspond to $\varepsilon$ frequency points. The more the number of frequency points that are sampled, the better the signal processing gain can be controlled. However, direct formation of the constraint matrix $\mathbf{C}$ from the constraint vectors corresponding to the numerous frequency points is not the most efficient approach because its computational complexity can increase significantly.

For this reason, we utilize the eigenvector constraint design scheme [30], which is a better approach that is based on the low-rank representation of signals. It is guaranteed that this approach minimizes the mean square error between the desired and actual beamformer responses over the frequency region of interest, with respect to a given number of constraints [43]. Therefore, we incorporated this design scheme with the received signal model of the dual-polarized antenna derived in Section II-B to attain effectiveness of the proposed constraint matrix approach and reduce its computational complexity. It should be noted that, to the best of our knowledge, no study has applied the eigenvector constraint design scheme for GPS antijam signal processing.

The first step of the proposed approach is to form a constraint matrix $\mathbf{C}$ using $\varepsilon$ constraint vectors that are obtained by sampling the frequency band of interest uniformly with $\varepsilon$ frequency points (that is, $f_1, f_2, \ldots, f_\varepsilon$), and $\varepsilon$ is usually chosen to be a large number. Then, with respect to the incoming GPS signal from $(\varphi_{GPS}, \theta_{GPS})$ direction, the novel constraint matrix $\mathbf{C}$ of STPAPS is formulated using the vector $\tilde{\mathbf{c}}(\varphi, \theta, \gamma, \eta, f)$ in (11) as follows:

$$\begin{aligned} \mathbf{C}_{2M \times \varepsilon} = [&\tilde{\mathbf{c}}(\varphi_{GPS}, \theta_{GPS}, 45°, -90°, f_1) \\ &\cdots \tilde{\mathbf{c}}(\varphi_{GPS}, \theta_{GPS}, 45°, -90°, f_\varepsilon)] \end{aligned} \quad (21)$$

In our experiments in Section IV, the GPS signal direction ($\varphi_{GPS}$, $\theta_{GPS}$) was obtained from the GPS ephemerides in the navigation data bits and the initial position fix of the receiver. We assumed that the initial GPS position fix of the receiver is available before the injection of jamming signals. If the receiver is turned on under a jamming situation, the initial GPS position fix may not be available. In that case, the conventional power minimization antijamming technique [25] that does not require the direction of GPS signals can be used to obtain the initial position of the receiver. Since the power minimization method does not perform beamforming, its antijamming performance is expected to be worse than the performance of beamforming methods [22]. Nevertheless, the power minimization is useful to initiate beamforming methods when the GPS signal direction is difficult to obtain owing to the lack of the initial position estimate. For a dynamic vehicle, the attitude information of the vehicle is also required to obtain the direction ($\varphi_{GPS}$, $\theta_{GPS}$), similar to that in the case of conventional MVDR-based GPS CRPAs, because ($\varphi_{GPS}$, $\theta_{GPS}$) is relative to the vehicle. As shown in (21), each column vector of $\mathbf{C}$ is equal to the constraint vector in (11) and is used to maintain the gain in (10) of the received RHCP (i.e., $\gamma = 45°$, $\eta = -90°$) GPS signal from the ($\varphi_{\text{GPS}}$, $\theta_{\text{GPS}}$) direction to unity, with respect to the corresponding frequency point. In addition, as desired, all the filter time taps (i.e., $2M$ time taps in Fig. 2) are involved in the formation of the proposed constraint vector $\tilde{\mathbf{c}}(\varphi_{\text{GPS}}, \theta_{\text{GPS}}, 45°, -90°, f_\rho)$ ($\rho = 1, 2, \ldots, \varepsilon$), as shown in (11) and (12). From the difference between the constraint vectors of (11) and (15) (for STPAPS and the previous dual-polarized antenna methods, respectively), it can be seen that unlike (15), there are no zero terms in (11) because all the time taps are constrained.

The second step is to factorize $\mathbf{C}$ into the product of three matrices by the singular value decomposition (SVD), as follows:

$$\mathbf{C} = \mathbf{U}\boldsymbol{\Sigma}\mathbf{V}^H \tag{22}$$

where $\mathbf{U}$ is a $2M \times 2M$ unitary matrix, $\boldsymbol{\Sigma}$ is a $2M \times \varepsilon$ rectangular diagonal matrix that contains the singular values of $\mathbf{C}$ in descending order, and $\mathbf{V}$ is an $\varepsilon \times \varepsilon$ unitary matrix.

In the third step, which is the essential part of the eigenvector constraint design, it is required to find an approximated constraint matrix $\mathbf{C}_\zeta$ of rank $\zeta$ ($\zeta \ll \varepsilon$), with respect to $\mathbf{C}$. This procedure is based on minimizing the Frobenius norm of the difference between the two matrices $\mathbf{C}_\zeta$ and $\mathbf{C}$ [44]. After splitting the three matrices $\mathbf{U}$, $\boldsymbol{\Sigma}$, and $\mathbf{V}$ with respect to the given number $\zeta$, $\mathbf{C}_\zeta$ is obtained as follows:

$$\mathbf{U} = \begin{bmatrix} \mathbf{U}_\zeta & \bar{\mathbf{U}}_\zeta \end{bmatrix}, \boldsymbol{\Sigma} = \begin{bmatrix} \boldsymbol{\Sigma}_\zeta & \mathbf{0} \\ \mathbf{0} & \bar{\boldsymbol{\Sigma}}_\zeta \end{bmatrix}, \mathbf{V} = \begin{bmatrix} \mathbf{V}_\zeta & \bar{\mathbf{V}}_\zeta \end{bmatrix} \tag{23}$$

$$\mathbf{C}_\zeta = \mathbf{U}_\zeta \boldsymbol{\Sigma}_\zeta \mathbf{V}_\zeta^H$$

where $\mathbf{U}_\zeta$ and $\mathbf{V}_\zeta$ consist of the first $\zeta$ columns of $\mathbf{U}$ and $\mathbf{V}$, and $\boldsymbol{\Sigma}_\zeta$ is a diagonal matrix whose diagonal entries correspond to the first $\zeta$ largest singular values of $\boldsymbol{\Sigma}$. The remaining columns of $\mathbf{U}$ and $\mathbf{V}$ and the diagonal matrix with the remaining singular values of $\boldsymbol{\Sigma}$ are denoted by $\bar{\mathbf{U}}_\zeta$, $\bar{\mathbf{V}}_\zeta$, and $\bar{\boldsymbol{\Sigma}}_\zeta$, respectively.

As shown in (23), the selection of $\zeta$ is a key part of this step. In [30], it was shown that the number of bases, $D$, needed to span the signal space over the frequency band [$f_{\min}$, $f_{\max}$] is lower bounded as follows, given the bandwidth $B$ of the complex-valued signal from $f_{\min}$ to $f_{\max}$:

$$\begin{aligned} B &= \pi(f_{\max} - f_{\min}) \\ T &= \tau_{N-1} + (M-1)T_s, \\ D &\geq \left\lceil \frac{BT}{\pi} + 1 \right\rceil \end{aligned} \tag{24}$$

where $T$ is the temporal duration required for the signal to propagate through the beamformer to the output from the time it first reaches the array, $\tau_{N-1}$ is the propagation delay that occurs when the incoming signal propagates from the reference antenna to the $N$-th antenna located furthest from the reference antenna [44], $T_s = 1/f_s$ is the sampling time, and $\lceil \chi \rceil$ is the ceiling function of $\chi$ which returns the smallest integer that is not smaller than $\chi$. To find the approximated constraint matrix $\mathbf{C}_\zeta$, we choose the value $\zeta$ as the minimum value of $D$.

Once $\mathbf{C}_\zeta$ has been obtained, the matrix $\mathbf{C}$ used for the constraint equation in (20) is replaced with $\mathbf{C}_\zeta$, and then (20) is changed as follows:

$$\min_{\mathbf{w}} \mathbf{w}^H \mathbf{R} \mathbf{w} \text{ subject to } \mathbf{w}^H \mathbf{C}_\zeta = \mathbf{1}_\varepsilon \tag{25}$$

where the constraint equation in (25) can be rewritten as follows using (23):

$$\mathbf{w}^H \mathbf{U}_\zeta = \mathbf{1}_\varepsilon \mathbf{V}_\zeta \boldsymbol{\Sigma}_\zeta^{-1} \tag{26}$$

Using the constraint equation expressed in (26), we derived the optimal weight vector that satisfies (25) as follows using the Lagrange multiplier method:

$$\mathbf{w}_{\text{opt}} = \tilde{\mathbf{R}}^{-1} \mathbf{U}_\zeta (\mathbf{U}_\zeta^H \tilde{\mathbf{R}}^{-1} \mathbf{U}_\zeta)^{-1} (\mathbf{1}_\varepsilon \mathbf{V}_\zeta \boldsymbol{\Sigma}_\zeta^{-1})^H \tag{27}$$

where $\tilde{\mathbf{R}}$ is an estimate of the covariance matrix $\mathbf{R}$ at the $k$-th sampling epoch using a finite number of data samples, as follows:

$$\tilde{\mathbf{R}} = \frac{1}{L} \sum_{l=k-L+1}^{k} \mathbf{v}[l]\mathbf{v}^H[l] \tag{28}$$

where $L$ is number of samples used to calculate $\tilde{\mathbf{R}}$; one millisecond samples were used in our implementation (i.e., $L = 5000$). It should be noted that a singularity can occur if the number of data samples are not sufficient when estimating the covariance matrix [45]–[48]. In that case, the estimated covariance matrix has singular values of zero and the inverse of the estimated covariance matrix does not exist [45]–[48]. Therefore, it is important to have a sufficient number of data samples (e.g., 5000 samples as used in this work) for covariance estimation.

The other parameters specifically used for implementing STPAPS were as follows:

$$\varepsilon = 100, f_{\min} = -2\,\text{MHz}, f_{\max} = 2\,\text{MHz}, f_s = 5\,\text{MHz} \tag{29}$$

The values of $f_{\min}$ and $f_{\max}$ provide sufficient bandwidth for the GPS L1 C/A code signal, and $f_s$ is greater than the Nyquist rate. The $\varepsilon$ value was determined experimentally. With an increase in the number of frequency points, $\varepsilon$, up to about 100, $C/N_0$ tended to increase; however, no noticeable increase in $C/N_0$ could be observed with further increase in $\varepsilon$. Thus, a value of 100 for $\varepsilon$ was a good compromise between the performance and computational complexity. Besides, $\tau_{N-1}$ in (24) is equal to zero, because a single-element antenna is used in this study. The number of filter time taps for all the methods compared in this study was set to 15, that is, $M = 15$ in Fig. 2. Based on these parameters and (24), $\zeta$, which is the same as the minimum value of $D$, was calculated to be 13.

IV. EXPERIMENTAL RESULTS AND DISCUSSIONS

To demonstrate the antijamming capability of STPAPS, we performed experiments under various jamming conditions in both low and high multipath environments. The performance of STPAPS was compared with that of 1) other MMSE- or MVDR-based methods for dual-polarized antennas in the literature, 2) other single-element dual-polarized antenna methods that are not based on MMSE or MVDR algorithms, and 3) the conventional MVDR-based STAP algorithm with a two-element single-polarized antenna array because both methods have one DOF.

A. Hardware setup

In order to obtain live GPS signals, two identical dual-polarized antennas (produced by Antcom) were installed on the rooftop of a building at Yonsei University, Incheon, Korea. As shown in Fig. 3, the interspacing was set to be half the wavelength of the GPS L1 signal, that is, 9.5 cm. The first antenna (Antenna 1 in Fig. 3) was located at the origin of the coordinates in Fig. 1, and the second antenna (Antenna 2 in Fig. 3) was placed on the $x$-axis, 9.5 cm away from the origin. These antennas were installed on the rooftop in both low and high multipath environments. To create the high multipath environment, the antennas were located close to a tall wall structure on the rooftop.

The RHCP and LHCP signals received by Antenna 1 were used as input data for STPAPS and the other dual-polarized antenna algorithms [22], [26] as illustrated in Fig. 4. In contrast, only the RHCP signals received by Antenna 1 and Antenna 2 were used as the input data for the conventional two-element single-polarized antenna array algorithm. (Please note that STPAPS was compared with the previous methods in three categories as mentioned at the beginning of Section IV.) The three RF signals (two RHCP signals and one LHCP signal) received by the two dual-polarized antennas were recorded as digital IF data by the hardware setup, built as depicted in Fig. 4. For down-converting and sampling the RF signals, Universal Software Radio Peripherals (USRPs) were used. The three USRPs were synchronized using Octoclock-g from Ettus Research, which could generate 10 MHz and 1 PPS clock signals using a GPS-disciplined oscillator. The local oscillator frequency and the sampling frequency of the USRPs were set to be 1575.42 MHz (i.e., down-converted to the baseband) and 5 Msps (i.e., $5 \times 10^6$ inphase and $5 \times 10^6$ quadrature samples per second), respectively. Finally, the sampled signals were saved as 32-bit floating-point data.

### B. Complex radiation field functions of the dual-polarized antenna

The complex radiation field functions, $\mathbf{G}_R(\varphi,\theta)$ and $\mathbf{G}_L(\varphi,\theta)$, are essential for obtaining the constraint matrix in (21). Similar to (3), $\mathbf{G}_R(\varphi,\theta)$ and $\mathbf{G}_L(\varphi,\theta)$ can be expressed as follows:

$$\begin{aligned}\mathbf{G}_R(\varphi,\theta) &= G_{R,\varphi}(\varphi,\theta)\exp\left[jP_{R,\varphi}(\varphi,\theta)\right]\hat{\mathbf{e}}_\varphi \\ &\quad + G_{R,\theta}(\varphi,\theta)\exp\left[jP_{R,\theta}(\varphi,\theta)\right]\hat{\mathbf{e}}_\theta \\ \mathbf{G}_L(\varphi,\theta) &= G_{L,\varphi}(\varphi,\theta)\exp\left[jP_{L,\varphi}(\varphi,\theta)\right]\hat{\mathbf{e}}_\varphi \\ &\quad + G_{L,\theta}(\varphi,\theta)\exp\left[jP_{L,\theta}(\varphi,\theta)\right]\hat{\mathbf{e}}_\theta\end{aligned} \quad (30)$$

To obtain the eight gain and phase patterns of the dual-polarized antenna, namely $\{G_{R,\varphi}(\varphi,\theta), G_{R,\theta}(\varphi,\theta), G_{L,\varphi}(\varphi,\theta), G_{L,\theta}(\varphi,\theta)\}$ and $\{P_{R,\varphi}(\varphi,\theta), P_{R,\theta}(\varphi,\theta), P_{L,\varphi}(\varphi,\theta), P_{L,\theta}(\varphi,\theta)\}$, far field measurements were performed in an anechoic chamber at the GPS L1 frequency (1575.42 MHz). One dual-polarized antenna that could receive RHCP and LHCP signals was placed inside the chamber and another dual-polarized antenna that could transmit horizontally and vertically polarized signals was used as the source antenna. The positioner rotated the antenna under test relative to the source antenna, and the patterns were measured every 3° in both the $\varphi$ and $\theta$ directions over the entire upper hemisphere region in Fig. 1, that is, $3° \leq \varphi \leq 360°$ and $0° \leq \theta \leq 90°$. As a result, all the eight gain and phase pattern data were acquired in the form of 120×31 two-dimensional grid-point values. Utilizing these data, the same-size grid-point values of $\mathbf{G}_R(\varphi,\theta)$ and $\mathbf{G}_L(\varphi,\theta)$ were calculated. Then, the values of $\mathbf{G}_R(\varphi,\theta)$ and $\mathbf{G}_L(\varphi,\theta)$ in an arbitrary direction $(\varphi, \theta)$ were obtained by performing conventional bilinear interpolation using four adjacent grid-point values.

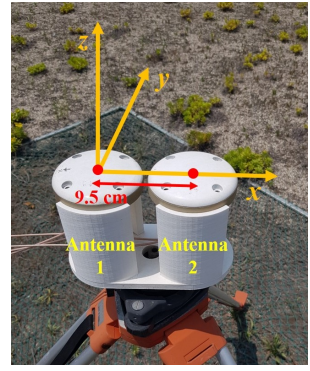

Fig. 3. Two dual-polarized antennas installed on the rooftop for the experiments.

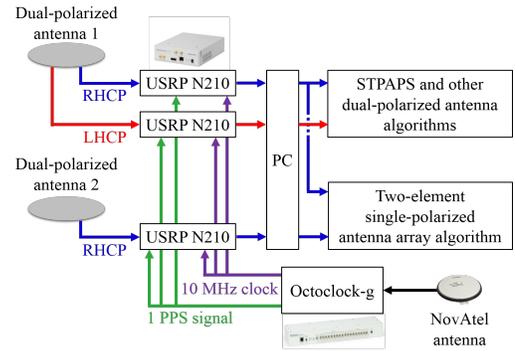

Fig. 4. Hardware setup for recording the received signals.

In Fig. 5, the gain and phase patterns of the RHCP antenna component of the dual-polarized antenna, which were measured with respect to the vertically polarized incoming signals (i.e., $G_{R,\theta}(\varphi,\theta)$ and $P_{R,\theta}(\varphi,\theta)$ in (30), respectively), are shown as examples. The two patterns are depicted in polar coordinates. The radii of the equally spaced concentric circles correspond to $\theta$, i.e., the center point and the outermost circle correspond to $\theta = 0°$ and $\theta = 90°$, respectively. The polar angle corresponds to $\varphi$ (in degrees) in Fig. 1.

### C. Comparison with other dual-polarized antenna methods using MMSE or MVDR algorithms

The performance of STPAPS was compared with that of the existing dual-polarized antenna antijamming methods that are based on the MMSE algorithm [26] or the MVDR algorithm using a single-frequency constraint vector [22]. Although those methods were originally developed for dual-polarized antenna arrays, they are applicable to single-element dual-polarized antennas as well, which is the case when the number of array elements is one. The problem formulations and optimal weight vectors of those methods are summarized in Table I. There is another related journal publication [27], but the work of [27] utilized the same constraint vector as that of [22] and focused on improving its computational efficiency. Therefore, it is not included in Table I.

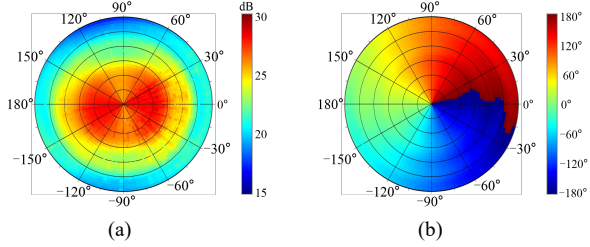

Fig. 5. (a) Gain and (b) phase patterns of the RHCP antenna component of the dual-polarized antenna measured with respect to the vertically polarized incoming signals.

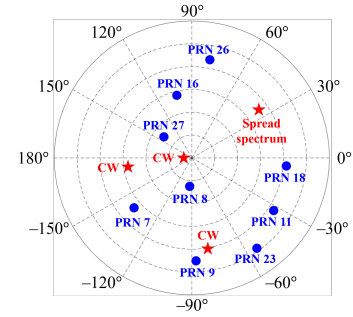

Fig. 6. Sky plot of nine GPS satellites and four jammers (Table II) for the experiment.

The jamming scenario in Table II represents a typical situation with one spread spectrum and three CW synthetic jammers. (The sky plot of GPS satellites and jammers is given in Fig. 6) The synthetic jamming signals were injected into the recorded live GPS signals by post-processing. The radiation patterns of the antenna were considered for the synthetic jammer injection. The jammers had arbitrarily chosen directions, polarizations, and center frequencies. To generate the spread spectrum jammer, a sequence of chips with an amplitude of +1 or –1 was modulated by the carrier signal at the GPS L1 frequency. The chipping rate of the sequence was set to 2.5575 Mcps, in which case the null-to-null bandwidth of the jammer became approximately 5 MHz, which covers the entire receiver frequency range, i.e., [$-f_s/2 = -2.5$ MHz, $f_s/2 = 2.5$ MHz]. This type of jammer is sometimes called wideband jammer or broadband jammer in the related GPS antijamming papers [49]–[53]. The CW jammers were continuous sinusoidal waves at the given frequencies. The jammer-to-noise-power ratios (JNRs) of the jammers were set to 40 dB and the jammers were injected at 130 ms from the beginning of each experiment.

The resultant $C/N_0$ plots of the software defined receiver (SDR) channel tracking GPS PRN 8 signals in low and high multipath environments are depicted in Figs. 7(a) and 7(b), respectively. Each $C/N_0$ value was calculated every 20 ms (i.e., for every navigation data bit). The red, green, and blue lines in

TABLE I
DUAL-POLARIZED ANTENNA-BASED GPS ANTIJAMMING ALGORITHMS

| Reference | Type of processing | Problem formulation for adaptive antijam processing | Optimal weight vector |
|---|---|---|---|
| Fante [26] | MMSE | $\min_{\mathbf{w}} E\{u_d[k]-u[k]\}$ <br> $u_d[k]$: locally generated GPS signal | $\mathbf{w}_{opt} = \tilde{\mathbf{R}}^{-1}\mathbf{p}^*$ <br> $\mathbf{p} = E\{\mathbf{v}[k]u_d^*[k]\}$ |
| Park et al. [31] | MVDR using a single-frequency constraint vector | $\min_{\mathbf{w}} \mathbf{w}^H \mathbf{R} \mathbf{w}$ subject to $\mathbf{w}^H \mathbf{c} = 1$ <br> $\mathbf{c} = \begin{bmatrix} c_R & c_L & \mathbf{0}_{1 \times 2(M-1)} \end{bmatrix}^T$ <br> $c_R = \mathbf{G}_R(\varphi_{GPS}, \theta_{GPS}) \cdot \hat{\mathbf{E}}(\varphi_{GPS}, \theta_{GPS}, 45°, -90°)^*$ <br> $c_L = \mathbf{G}_L(\varphi_{GPS}, \theta_{GPS}) \cdot \hat{\mathbf{E}}(\varphi_{GPS}, \theta_{GPS}, 45°, -90°)^*$ | $\mathbf{w}_{opt} = \dfrac{\tilde{\mathbf{R}}^{-1}\mathbf{c}}{\mathbf{c}^H \tilde{\mathbf{R}}^{-1}\mathbf{c}}$ |
| Proposed | MVDR using the eigenvector constraint design scheme | $\min_{\mathbf{w}} \mathbf{w}^H \mathbf{R} \mathbf{w}$ subject to $\mathbf{w}^H \mathbf{C}_\zeta = \mathbf{1}_\varepsilon$ | $\mathbf{w}_{opt} = \tilde{\mathbf{R}}^{-1}\mathbf{U}_\zeta$ <br> $\cdot (\mathbf{U}_\zeta^H \tilde{\mathbf{R}}^{-1} \mathbf{U}_\zeta)^{-1}$ <br> $\cdot (\mathbf{1}_\varepsilon \mathbf{V}_\zeta \mathbf{\Sigma}_\zeta^{-1})^H$ |

TABLE II
JAMMER CHARACTERISTICS

| Jammer type | Direction ($\varphi, \theta$) | Polarization ($\gamma, \eta$) | Intermediate center frequency |
|---|---|---|---|
| Spread spectrum | (35.86°, 54.08°) | (63.31°, 31.08°) | 0.217 MHz |
| CW | (179.64°, 5.29°) | (67.42°, 42.34°) | −2.207 MHz |
| CW | (−172.00°, 42.15°) | (34.15°, −95.58°) | 2.447 MHz |
| CW | (−80.05°, 60.60°) | (76.25°, 132.54°) | −2.359 MHz |

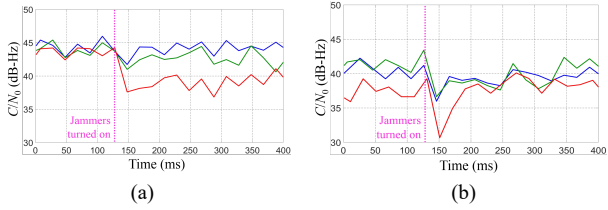

Fig. 7. $C/N_0$ comparison of the MMSE (red), MVDR using a single-frequency constraint vector (green), and STPAPS (blue) methods summarized in Table I under the influence of the jammers listed in Table II in (a) low and (b) high multipath environments. The four jamming signals were injected at 130 ms, and their JNRs were set to 40 dB.

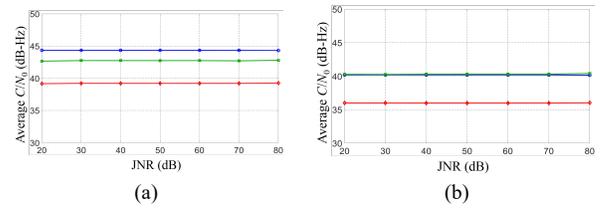

Fig. 8. Average $C/N_0$ comparison of the MMSE (red), MVDR using a single constraint vector (green), and STPAPS (blue) methods summarized in Table I under the influence of the jammers listed in Table II in (a) low and (b) high multipath environments.

Fig. 7 indicate the MMSE [26], MVDR using a single-frequency constraint vector [22], and STPAPS algorithms in Table I, respectively. Although the results presented here are with respect to a single GPS channel, similar results can be obtained for other PRNs because all the GPS signal tracking channels operate in parallel with the same algorithm.

When the jammers were injected at 130 ms in the low multipath environment, as shown in Fig. 7(a), the MVDR method using a single-frequency constraint vector [22] (green) showed a slight $C/N_0$ degradation while the MMSE method [26] (red) showed significant $C/N_0$ degradation. On the other hand, STPAPS (blue) did not show any such degradation. The $C/N_0$ drop with the MVDR method [22] is due to the undesired attenuation of the GPS signals during antijamming processing, which is a limitation of the single-frequency constraint vector considering the single time tap explained in Section III-B. The $C/N_0$ degradation of the MMSE method [26] is caused by a different reason. The performance of the MMSE algorithm degrades if the cross-correlation between desired and undesired signals is not negligible [54], which is the case with the GPS PRN 8 signals (desired) and the spread spectrum jammer (undesired) in Table II. STPAPS did not show any noticeable $C/N_0$ degradation when the jamming signals were incident, because it overcomes the limitation of [22] by utilizing the newly proposed constraint matrix, which constrains multiple frequency points and all the filter time taps. Note the difference between the previous constraint vector in (15) and the proposed constraint matrix in (18).

Similar experiments were performed using the same jammers but different JNRs, and the average $C/N_0$ values of the three methods are shown in Fig. 8(a). For each JNR, the $C/N_0$ of each method was averaged over the duration of the experiment after jammer injection (i.e., from 130 to 400 ms). It is clear that STPAPS outperforms both the MVDR and MMSE methods [22], [26] with respect to all JNRs. All three methods showed almost constant average $C/N_0$ values for all JNRs. This is consistent with the observation of previous studies that MVDR and MMSE beamformers show nearly constant performance in certain ranges of JNR (see [55] and [56]).

On the other hand, in the high multipath environment, all the three techniques experienced performance degradation in both the absence and presence of jamming signals, as shown in Figs. 7(b) and 8(b), when compared to that in the case of the low multipath environment shown in Figs. 7(a) and 8(a). The weight vectors computed by STPAPS and the MVDR using a single constraint vector [22] maintained the signal processing gain in the GPS signal direction. The GPS signal direction is almost the same as the GPS satellite direction under low multipath environment, and therefore it is usually calculated using the ephemeris data in the GPS signals. However, in the case of high multipath environment, although the line-of-sight (LOS) GPS signal direction is the same as the GPS satellite direction, the number and directions of the multipath signals are not readily known. Therefore, the application of a weight vector that is calculated only for the LOS signal to the multipath signals does not provide any $C/N_0$ benefit. The multipath signals in this case are treated as interference signals and thus partly mitigated, but they cannot be completely mitigated by STPAPS and other dual-polarized methods.

This performance degradation of MVDR-based beamformers in the presence of multipaths was also observed in previous studies [57]–[60], and the $C/N_0$ drop due to multipaths is commonly observed when using conventional GPS receivers [61]–[64]. It should be noted that the constraint matrix approach of STPAPS did not show any $C/N_0$ improvement over the conventional single-frequency constraint vector approach under the high multipath environment. This is because both approaches constrain the weights only for the LOS signals and cannot completely mitigate the multipath signals. The additional performance degradation of the MMSE method [26] due to multipaths is understandable because the correlations between the LOS and multipath signals is not negligible in a high multipath environment.

### D. Comparison with other single-element dual-polarized antenna methods

A qualitative comparison of STPAPS and the other single-element dual-polarized antenna techniques that are not based on the MMSE or MVDR algorithms is summarized in Table III. It was not possible to quantitively evaluate the performance of each method because the special antenna hardware implemented by [35] and [37] was not available to us, and the exact algorithm of [36] was not revealed in the paper. From Table III, it can be seen that no single method is superior to all the others. The advantage of STPAPS is that it can mitigate an RHCP spread spectrum jamming signal, which has not been demonstrated in [35] or [36]. In addition, multiple jammers with different polarizations can be mitigated by STPAPS. Since STPAPS is concerned with jamming signals originating from above the horizon, this method can complement the technique proposed in [37], which overcomes a jammer originating from below the horizon. However, in contrast to the method presented in [35], STPAPS cannot mitigate multiple wideband jamming signals.

TABLE III
COMPARISON OF SINGLE-ELEMENT DUAL-POLARIZED ANTENNA BASED GPS ANTIJAMMING METHODS

| Method | Number and type of mitigatable jammer | Polarization of mitigatable jammer | Reported mitigation performance |
|---|---|---|---|
| Rosen and Braasch [35] | Single wideband Gaussian noise and single narrowband, or multiple wideband if their polarizations are identical | Other than direct RHCP | 20–30 dB interference suppression in the presence of a wideband jammer with 30 dB jammer-to-noise-power ratio (JNR) |
| Kraus et al. [36] | Single chirp or single filtered noise with 1 MHz 3-dB bandwidth | Only LP | $C/N_0$ is maintained between 40 and 45 dB-Hz in the presence of a jammer with 30 dB jammer-to-signal-power ratio (JSR) |
| McMilin et al. [37] | Single spread spectrum (below the horizon) | Any polarizations | $C/N_0$ in the case of antijam mode is 15 dB-Hz higher than $C/N_0$ in the case of normal mode |
| STPAPS | Single spread spectrum with 5 MHz null-to-null bandwidth and multiple CW with various polarizations (above the horizon) | Any polarizations | $C/N_0$ is maintained between 40 and 45 dB-Hz in the presence of a spread spectrum jammer with 30 dB JSR |

Throughout the experiment, STPAPS maintained a $C/N_0$ of nearly 44 dB-Hz even in the presence of jammers with 40 dB JNR as shown in Fig. 7(a). This performance is comparable with the antijamming performance reported in [35] and [36]. The GPS SDR used in this study can track signals if $C/N_0$ is 30 dB-Hz or above. Under the jamming scenario of this experiment, the GPS SDR could not track the GPS signals without turning on the antijam capability, which means that $C/N_0$ was smaller than 30 dB-Hz under jamming. Thus, the $C/N_0$ benefit resulting from the use of STPAPS is more than 14 dB-Hz. This $C/N_0$ benefit is comparable with that of [37]. Although the $C/N_0$ benefit from STPAPS is similar to that in previous studies, our algorithm has a unique strength that it can mitigate multiple jammers above the horizon with any polarizations.

### E. Comparison with the conventional two-element single-polarized antenna array algorithm

Here, we compare the characteristics of STPAPS and the conventional two-element single-polarized antenna array STAP. This is worth comparing because both methods have one DOF, and a detailed comparison between single-element dual-polarized antenna and two-element single-polarized antenna array methods is not available in the literature. We considered the four special cases illustrated in Fig. 9, which elucidate the different characteristics of the two antijamming methods. A single spread spectrum jammer with 40 dB JNR, which was generated in the same way as explained in Section IV-C, having a specific direction and polarization was injected under each special jamming case. Detailed information about the jammer used in each case is given in Table IV.

The JNRs of the jammers were set to 40 dB and each jammer was injected at 130 ms from the beginning of each experiment. The resultant $C/N_0$ of the GPS PRN 8 signals are shown in Fig. 10, where the red and blue curves indicate the conventional two-element array STAP and STPAPS, respectively.

*1) Special case 1: A non-RHCP jammer and a GPS signal in the same direction*

In special case 1, a spread spectrum jammer was incident from a direction same as that of the GPS signal as shown in Fig. 9, that is, $(\varphi_{SSJ}, \theta_{SSJ}) = (\varphi_{GPS}, \theta_{GPS}) = (-94.29°, 18.75°)$, where $(\varphi_{SSJ}, \theta_{SSJ})$ is the direction of the jamming signal. The polarization $(\gamma, \eta)$ of the jammer was set as $(42°, -87°)$ so that it is similar to the GPS signal polarization, which is $(45°, -90°)$, to observe its influence.

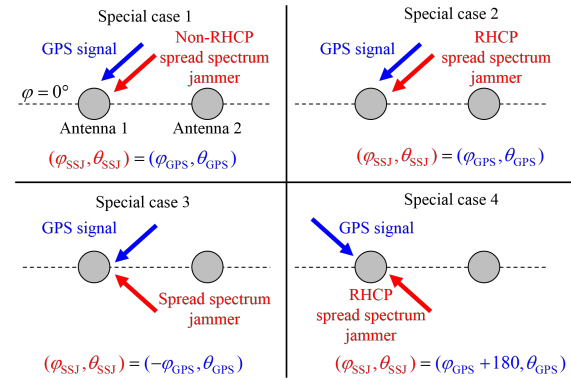

Fig. 9. Conceptual illustration of the jamming scenarios in the four special cases. STPAPS uses the RHCP and LHCP signals received by Antenna 1, while the conventional two-element single-polarized antenna array STAP uses the RHCP signals received by Antennas 1 and 2 (in Fig. 3). Both methods have one DOF, but the size of the proposed single-antenna solution (i.e., STPAPS) is significantly smaller than that of the antenna array solution.

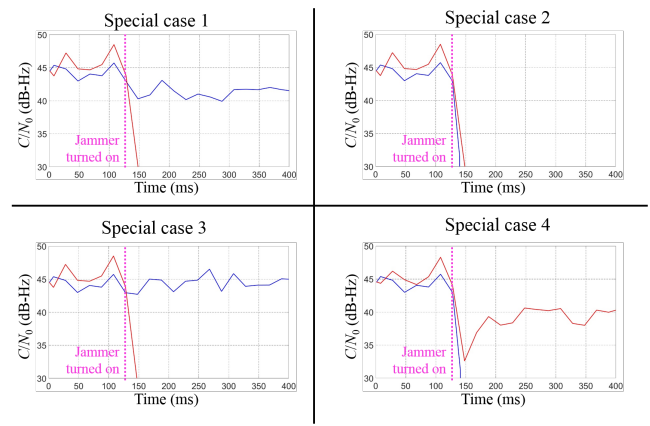

Fig. 10. $C/N_0$ comparison between the two-element single-polarized array STAP (red) and STPAPS (blue) under the four special jamming cases. Each special case is illustrated in Fig. 9.

The conventional STAP algorithm using a single-polarized antenna array generates a spatial null (i.e., low antenna array gain) toward the jammer while maintaining the gain in the GPS direction by using the constraint vector given by (18). However, if the jamming and GPS signals are from the same direction, the conventional STAP method cannot generate a null in the jammer direction because the constraint vector constrains the gain in that direction. Thus, the conventional STAP algorithm loses its lock on the GPS signal as indicated by the red curve of special case 1 in Fig. 10.

TABLE IV
JAMMER CHARACTERISTICS OF EACH SPECIAL CASE SHOWN IN FIG. 9

| Case | Jammer type | Direction ($\varphi, \theta$) | Polarization ($\gamma, \eta$) | Intermediate center frequency |
|---|---|---|---|---|
| Special case 1 | Spread spectrum | (−94.29°, 18.75°) | (42°, −87°) | 1.461 MHz |
| Special case 2 | Spread spectrum | (−94.29°, 18.75°) | (45°, −90°) | 0.276 MHz |
| Special case 3 | Spread spectrum | (94.29, 18.75°) | (74°, −139°) | 0.732 MHz |
| Special case 4 | Spread spectrum | (85.71°, 18.75°) | (45°, −90°) | 0.510 MHz |

In contrast, STPAPS, which utilizes polarization diversity, can mitigate the jammer in both the polarization domain and spatial domain. Hence, if the polarizations of the jamming and GPS signals are not identical (i.e., if the polarization of the jamming signal is non-RHCP), STPAPS can reject only the jammer while preserving the GPS signals even when both signals are from the same direction. This was experimentally validated as indicated by the blue curve of special case 1 in Fig. 10.

It should be noted that the $C/N_0$ of STPAPS degrades after the jammer injection at 130 ms (blue curve of special case 1 in Fig. 10). This is due to the similarity in polarization ($\gamma, \eta$) of the jamming and GPS signals, which were (42°, −87°) and (45°, −90°), respectively. During the jammer mitigation in the polarization domain, the GPS signals are also partly mitigated if their polarizations are similar. This phenomenon is analogous to the $C/N_0$ degradation of GPS signals with the conventional STAP algorithm when the jammer direction is similar to the GPS signal direction. When the polarization of the jammer was significantly different from that of the GPS signal (special case 3 in Fig. 10 and Table IV), STPAPS did not experience any $C/N_0$ degradation. Special case 3 will be discussed in detail in a later subsection.

*2) Special case 2: An RHCP jammer and a GPS signal in the same direction*

If the polarization of the jammer is RHCP and the jamming and GPS signals are incident from the same direction, even the polarization diversity that STPAPS utilizes cannot distinguish and mitigate the jammer signal alone. Since both jamming and GPS signals had the same polarization and direction for special case 2 (as shown in Fig. 9), both STPAPS and conventional STAP could not mitigate the jammer as shown in Fig. 10 (for special case 2).

*3) Special case 3: An arbitrarily polarized jammer and a GPS signal with opposite azimuth angles*

In special case 3, the directions of an arbitrarily polarized jammer and the GPS signal were symmetric with respect to the $\varphi = 0°$ (or $y = 0$) plane, which contains a line passing through the two antenna locations, as illustrated in Fig. 9, i.e., ($\varphi_{SSJ}, \theta_{SSJ}$) = (−$\varphi_{GPS}, \theta_{GPS}$). In this case, the phase difference, i.e., $\Delta\psi_{GPS,2}$ in (19), between the GPS signals received by Antennas 1 and 2 is identical to that of the jamming signal due to the symmetry. ($\Delta\psi_{GPS,1} = 0$ because Antenna 1 is the reference antenna, and $N = 2$ in (19) for the two-element array.) Therefore, the constraint vector of conventional STAP using a two-element array to form a null toward the jammer, which is a function of $\Delta\psi_{GPS,2}$ as in (18) and (19), cannot distinguish between the jammer and GPS signal directions. Consequently, it failed to mitigate the jammer signal as shown by the red curve of special case 3 in Fig. 10.

On the other hand, the constraint matrix **C** of STPAPS given by (21) differs with respect to the two directions, because $\mathbf{G}_R(\varphi,\theta)$ and $\mathbf{G}_L(\varphi,\theta)$ in (30) are different from $\mathbf{G}_R(-\varphi,\theta)$ and $\mathbf{G}_L(-\varphi,\theta)$, respectively, and therefore $\tilde{\mathbf{c}}(\varphi,\theta,\gamma,\eta,f)$ in (11) differs from $\tilde{\mathbf{c}}(-\varphi,\theta,\gamma,\eta,f)$ regardless of signal polarization ($\gamma, \eta$). It was found that each gain pattern measured in this work had 180-degree symmetry in the azimuth ($\varphi$) direction, and each ph ase pattern increased by approximately 180° for each half revolution of $\varphi$ as shown in Fig. 5. For example, the phases at 60° and −120° azimuth in Fig. 5(b) are approximately 130° and −50°, respectively, which shows 180° phase difference. Hence, the complex radiation field functions of the dual-polarized antenna in this work have the following properties:

$$\mathbf{G}_R(\varphi+180°,\theta) \approx \mathbf{G}_R(\varphi,\theta)\exp[j180°]$$
$$\mathbf{G}_L(\varphi+180°,\theta) \approx \mathbf{G}_L(\varphi,\theta)\exp[j180°] \quad (31)$$

Consequently, STPAPS successfully mitigated the jammer in this case as demonstrated by the blue curve of special case 3 in Fig. 10. However, there exists a different kind of symmetry and ambiguity in (31), which leads to another interesting observation that will be discussed in the following subsection.

*4) Special case 4: An RHCP jammer and a GPS signal whose azimuth angles differ by 180 degrees*

The magnitudes of the complex radiation field functions in (31) would be almost identical if the azimuth angles of the two signals are separated by 180°. Special case 4 in Fig. 9 considers this situation, where an RHCP jammer has an azimuth angle that differs from that of the GPS signal by 180°, that is, ($\varphi_{SSJ}, \theta_{SSJ}$) = ($\varphi_{GPS} + 180°, \theta_{GPS}$). In this case, the constraint vector $\tilde{\mathbf{c}}(\varphi,\theta,\gamma,\eta,f)$ of STPAPS given by (11), which is obtained using the complex radiation field functions $\mathbf{G}_R(-\varphi,\theta)$ and $\mathbf{G}_L(-\varphi,\theta)$, also has the following property, because the polarizations ($\gamma, \eta$) of two incoming signals are identical.

$$\tilde{\mathbf{c}}(\varphi+180°,\theta,\gamma,\eta,f) \approx \tilde{\mathbf{c}}(\varphi,\theta,\gamma,\eta,f)\exp[j180°] \quad (32)$$

From (10) and (32) the following can be obtained:

$$\begin{aligned} g(\varphi+180°,\theta,\gamma,\eta,f) &= \mathbf{w}^H \tilde{\mathbf{c}}(\varphi+180°,\theta,\gamma,\eta,f) \\ &\approx \mathbf{w}^H \tilde{\mathbf{c}}(\varphi,\theta,\gamma,\eta,f)\exp[j180°] \\ &= g(\varphi,\theta,\gamma,\eta,f)\exp[j180°] \end{aligned} \quad (33)$$

As shown in (33), the signal processing gains with respect to the two directions $(\varphi, \theta)$ and $(\varphi + 180°, \theta)$ have almost the same magnitude. Therefore, STPAPS cannot reject signals from only one of those two directions.

In contrast, the conventional STAP method with a two-element array does not suffer from this ambiguity, because $\Delta\psi_{GPS,2}$ in (19), which is used for building the constraint vector in (18) for the two directions $(\varphi_{GPS}, \theta_{GPS})$ and $(\varphi_{GPS} + 180°, \theta_{GPS})$, is different. For these reasons, as shown in Fig. 10 (special case 4), STPAPS (indicated by the blue curve) lost the lock on the GPS signal, while the conventional STAP (indicated by the red curve) was able to continue tracking the GPS signal.

To summarize, the conventional STAP method has an ambiguity between the $(\varphi, \theta)$ and $(-\varphi, \theta)$ directions regardless of the polarizations of the two incoming signals (special case 3 in Figs. 9 and 10), whereas the $(\varphi, \theta)$ and $(\varphi + 180°, \theta)$ directions are ambiguous to STPAPS only if the two incoming signal polarizations are identical (special case 4 in Figs. 9 and 10).

### F. Comparison with the conventional two-element single-polarized antenna array algorithm

Table I lists the expressions of the optimum weight vector of each dual-polarized antenna method. The computational complexity of each method is governed by the weight vector calculation. First, let us consider the inversion of the covariance matrix $\tilde{\mathbf{R}}$ having a dimension of $2M \times 2M$. The computation of $\tilde{\mathbf{R}}^{-1}$ by using the conventional Gauss-Jordan elimination method is known to have a complexity of $O(M^3)$ [65].

The multiplication of a $2M \times 2M$ matrix $\tilde{\mathbf{R}}^{-1}$ and a $2M \times 1$ vector $\mathbf{p}^*$, which is required for [26], involves the complexity of $O(M^2)$. Thus, the overall complexity of [26] is $O(M^3)$. The method in [22] requires the multiplication of a $2M \times 2M$ matrix $\tilde{\mathbf{R}}^{-1}$ and $2M \times 1$ vector $\mathbf{c}$, but $\mathbf{c}$ contains only two non-zero terms. Therefore, this multiplication has a complexity of $O(M)$. This method also involves another multiplication, that is, of a $1 \times 2M$ vector $\mathbf{c}^H$ with two non-zero terms and a $2M \times 1$ vector $\tilde{\mathbf{R}}^{-1}\mathbf{c}$, which has a complexity of $O(1)$. Therefore, the overall complexity of [22] is also $O(M^3)$. STPAPS requires the SVD of a $2M \times \varepsilon$ constraint matrix $\mathbf{C}$, which has a complexity of $O(\varepsilon^2 M + \varepsilon M^2)$ when using the conventional bidiagonalization and QR algorithm [65]. STPAPS also requires the multiplication of a $2M \times 2M$ matrix $\tilde{\mathbf{R}}^{-1}$ and a $2M \times \zeta$ matrix $\mathbf{U}_\zeta$ (complexity of $O(M^2\zeta)$); multiplication of a $\zeta \times 2M$ matrix $\mathbf{U}_\zeta^H$ and a $2M \times \zeta$ matrix $\tilde{\mathbf{R}}^{-1}\mathbf{U}_\zeta$ (complexity of $O(M^2\zeta)$); and multiplications of an $1 \times \varepsilon$ vector $\mathbf{1}_\varepsilon$, an $\varepsilon \times \zeta$ matrix $\mathbf{V}_\zeta$, and a $\zeta \times \zeta$ diagonal matrix $\mathbf{\Sigma}^{-1}$ (complexity of $O(\varepsilon\zeta)$). Since $\mathbf{U}_\zeta^H \tilde{\mathbf{R}}^{-1}\mathbf{U}_\zeta$ is a $\zeta \times \zeta$ matrix, its inverse has a complexity $O(\zeta^3)$. Note that $M = 15$, $\varepsilon = 100$, and $\zeta = 13$ were used for STPAPS. Since $\varepsilon$ is usually greater than $M$ and $\zeta$, the overall complexity of STPAPS is $O(\varepsilon^2 M)$, which is dominated by the SVD calculation. To summarize, STPAPS is an $O(\varepsilon^2 M)$ algorithm while the other two methods [22], [26] are $O(M^3)$ algorithms.

In addition to the complexity analysis, the CPU time is usually used to compare the computational efficiency [66]–[68]. Using MATLAB 2020a, the CPU time required for calculating the weight vector of each algorithm in Table I was measured on a PC with an Intel Core i7-3770 CPU operating at 3.40 GHz and 32 GB of memory. After averaging over 100 runs of the weight vector calculation, the resultant CPU times were compared as shown in Table V. STPAPS takes approximately twice as much CPU time as the other methods. This may be because MATLAB performs a more efficient SVD calculation than the conventional bidiagonalization and QR algorithm [65] considered in our complexity analysis.

TABLE V
CPU TIME COMPARISON OF THE THREE DUAL-POLARIZED ANTENNA TECHNIQUES LISTED IN TABLE I

|  | Fante [26] | Park et al. [22] | STPAPS |
|---|---|---|---|
| CPU time (ms) | 0.792 | 0.817 | 1.629 |

## V. CONCLUSIONS

A novel GPS antijamming technique based on single-element dual-polarized antenna (i.e., STPAPS) was proposed and experimentally validated in this study. We derived the mathematical models of arbitrarily polarized signals received by the dual-polarized antenna and proposed an appropriate constraint matrix, which has not been proposed previously for a dual-polarized antenna. To reduce the computational complexity of the constraint matrix approach, the eigenvector constraint design scheme was first adopted for performing GPS antijam. We experimentally demonstrated and thoroughly compared the performance of STPAPS and the previous methods in three categories.

STPAPS provides higher $C/N_0$ than the existing MMSE or MVDR methods; however, its computational cost (measured in terms of CPU time) was approximately twice higher. In the low-multipath environment, approximately 5 dB or 2 dB $C/N_0$ improvement was obtained over the existing MMSE or MVDR methods, respectively. In the high-multipath environment, STPAPS and the existing MVDR method demonstrated a similar $C/N_0$, but both methods provided approximately 4 dB $C/N_0$ improvement over the existing MMSE method.

Compared to the other methods that are not based on the MMSE or MVDR algorithms, STPAPS has the advantage of mitigating a single spread spectrum and multiple CW jammers with arbitrary and different polarizations. However, STPAPS cannot mitigate multiple spread spectrum jammers and jammers below the horizon. Compared to the conventional STAP for a single-polarized array, STPAPS can mitigate 1) a non-RHCP spread spectrum jammer even when the jamming and GPS signals are incident from the same direction, and 2) a spread spectrum jammer from the $(-\varphi_{GPS}, \theta_{GPS})$ direction, which overcomes certain limitations of STAP. However, STPAPS cannot mitigate an RHCP spread spectrum jammer from the $(\varphi_{GPS} + 180°, \theta_{GPS})$ direction.

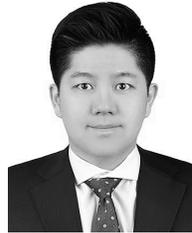

**Kwansik Park** is a Ph.D. candidate in the School of Integrated Technology, Yonsei University, Korea. He received his B.S. degree in electrical and electronic engineering from Yonsei University in 2012. His current research mainly focuses on GPS anti-interference technologies.

Mr. Park was a recipient of the Graduate Fellowship from the Information and Communications Technology (ICT) Consilience Creative Program supported by the Ministry of Science and ICT, Republic of Korea.

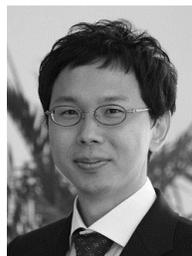

**Jiwon Seo** (M'12) received the B.S. degree in mechanical engineering (division of aerospace engineering) in 2002 from the Korea Advanced Institute of Science and Technology, Daejeon, Korea, and the M.S. degree in aeronautics and astronautics in 2004, the M.S. degree in electrical engineering in 2008, and the Ph.D. degree in aeronautics and astronautics in 2010 from Stanford University, Stanford, CA, USA.

He is currently an Associate Professor with the School of Integrated Technology, Yonsei University, Incheon, Korea. His research interests include complementary positioning, navigation, and timing systems, Global Navigation Satellite System (GNSS) antijamming technologies, and ionospheric effects on GNSS.

Prof. Seo is also a Member of the International Advisory Council of the Resilient Navigation and Timing Foundation, Alexandria, VA, USA, and a Member of several advisory committees of the Ministry of Oceans and Fisheries and the Ministry of Land, Infrastructure and Transport, Republic of Korea.